\documentclass[onecolumn,showpacs]{revtex4}
\usepackage{graphicx}
\usepackage{dcolumn}
\usepackage{bm}
\usepackage{amsmath}
\usepackage{amssymb}
\usepackage{epstopdf}
\usepackage{hyperref}

\begin{document}
{\setlength{\oddsidemargin}{1.2in}
\setlength{\evensidemargin}{1.2in} } \baselineskip 0.55cm
\begin{center}
{\LARGE {\bf FRW cosmology with varying cubic deceleration parameter}}
\end{center}
\date{\today}
\begin{center}
 Leishingam Kumrah, S. Surendra Singh, L. Anjana Devi and Md Khurshid Alam \\
   Department of Mathematics, National Institute of Technology Manipur,\\ Imphal-795004,India\\
   Email:{ lkumrah@gmail.com, ssuren.mu@gmail.com , anjnlam@gmail.com and alamkhurshid96@gmail.com}\\
 \end{center}

\textbf{Abstract}: In this work a new law of varying deceleration parameter of third degree have been proposed. The solutions of the modified field equations have been derived under the newly proposed law of the deceleration parameter. Model exhibits the Big-bang singularity at cosmic time ($t=0$) and shows Big Rip at ($t=n$) then it re-enter the phase of initial singularity at $t=2n$ and ends its cyclic behavior at $t=3n$. The evolution of the physical and dynamical parameters of the Universe have been studied and the graphical representation has also been shown. Further $Om(z)$ diagnostic parameter and the energy conditions have also been studied together with their graphical representations.\\


1. {\bf Introduction}\label{sec1}
\setcounter {equation}{0}
\renewcommand {\theequation}{\arabic{equation}}

From several observations of Riess et al. (1998, 1999), Perlmutter et al. (1999), Clocchiatti et al. (1999), CMB data of  Page (2003), WMAP of Spergel et al. (2003, 2007), Knop et al. (2003) \cite{1,2,3,4,5,6,7,8}, Planck collaborations of Ade et al. (2015) \cite{9}, it is shown that our Universe is undergoing an accelerated expansion. However, the cause behind the accelerating expansion of the Universe is not yet completely formulated. From existing literatures like Copeland et al. (2006) \cite{10}, Frieman et al. (2008) \cite{11} and Bamba et al. (2012) \cite{12}, we can say that dark energy plays a big role in the accelerating expansion of the Universe. But dark energy is still a mystery, although substantial lead has been formulated theoretically and observationally. Many cosmologists/researchers generally uses two different methods to explain the phenomenon of dark energy. One method is to adopt some exotic matter source such as quintessence (Martin 2008) \cite{13}, Chaplygin gas (Bento et al. 2002) \cite{14}, polytropic gas (Karami et al. 2009) \cite{15}, phantom models (Nojiri et al. 2003, 2009; Bilic et al. 2008), tachyons (Padmanabhan et al. 2002) \cite{16,17,18,19} and the cosmological constant $\Lambda$ (Alcaniz, 2006) \cite{20}. The second one is to modify the field equation of Einstein's general theory of relativity. The cosmological constant, which was introduced by Einstein in the field equation of general relativity is also considered as a candidate for dark energy (DE). However the cosmological constant term $\Lambda$ has suffered the so-called cosmological constant problem (Weinberg 1989 \cite{21} and Martin 2012 \cite{22}), fine-tuning and cosmic coincidence problem (Peebles and Ratra 2003 \cite{23}). Secondly, modifying the Einstein field equation, many modified theories of gravity has been obtained. Some notable examples are $f(R)$ (Sotiriou and Faraoni 2010 \cite{24}; De Felice and Tsujikawa 2010 \cite{25}), $f(R,G)$ (De Laurentis 2015 \cite{26}; Santos da Costa et al 2018 \cite{27}; Odintsov et al. 2019 \cite{28}; Singh 2021 \cite{29}), $f(R,T)$ (Harko et al. 2011 \cite{30}) and $f(Q,T)$ (Xu et al. 2019 \cite{31}) gravity theories.\\

Presently modified theories of gravity have become an interesting area of modern cosmology. Many researchers have developed different models of modified theory of gravities to accommodate the ever growing accelerating expansion of the Universe. In this setting $f(R)$ gravity, where $R$ is the Ricci scalar is one of the most simplest modification. It was first introduced by  Buchdahl (1970) \cite{32} and later used to find nonsingular isotropic de Sitter types cosmological solutions (Starobinsky 1980) \cite{33}. Many cosmologists/researchers have studied different features of $f(R)$  gravity in varied cosmological models (Nojiri and Odintsov 2007 \cite{34}; Carroll et al. 2004 \cite{35}; Reddy et al. 2014 \cite{36}). A more general modified cosmological model than $f(R)$ theory is the $f(R,T)$ gravity theory proposed by Harko et al. (2011) \cite{30}. It is established by coupling the geometry and matter in the gravitational action. Besides this $f(R)$ gravity has been expanded to $f(R,G)$ gravity, where $G$ is the Gauss-Bonnet scalar. It has been shown that $f(R,G)$ gravity obviously leads to an effective cosmological constant $\Lambda$, quintessence or phantom cosmic acceleration (Elizalde et al. 2010) \cite{37}. Xu et al. (2019) \cite{31} recently has proposed $f(Q,T)$ gravity, which is an extension of the symmetric teleparallel gravity. In this model, the gravitational action $L$ is represented by arbitrary function $f$ of the non-metricity $Q$ and trace of the energy-matter momentum tensor $T$. Simram and Sahoo (2020) \cite{38} investigated energy conditions in $f(Q,T)$ gravity and observes that the , weak, null, and dominant energy conditions are all satisfied while the strong energy conditions (sec) is violated as per the present accelerating expansion of the Universe.\\

Berman (1983) \cite{39}, Berman and Gomide (1988) \cite{40} had introduced a new law of time dependent Hubble parameter in the framework of Robertson-Walker space-time which yield constant deceleration parameter. Singh et al. (2009) \cite{41} investigated bulk viscous cosmological models of Universe in Lyra's Manifold by considering time varying deceleration parameter and coefficient of viscosity which is constant in FRW space-time, where exact solutions of Sen’s Equations in Lyra Geometry has been derived and shows that the Universe starts with a Big Bang singularity initially. Singh et al. (2010) \cite{42} studied a new class of bulk viscous cosmological models in a scale covariant theory of gravitation in which they have studied the false vacuum model, the stiff fluid model and radiating model with time varying deceleration parameter and found that the Universe has initial singularity. It also concludes that the Universe begins with Big Bang and is expanding. Singh (2015) \cite{43} investigated Friedmann Robertson Walker (FRW) Universe in the presence of viscous fluid based on Lyra’s manifold by considering linearly varying deceleration parameter and co-efficient of bulk viscosity to be a constant. Exact solutions has also been obtained where cosmological models have been derived. It has been found that the derived model of the Universe starts with initial singularity and have a finite lifetime, which ends with big rip. Akarsu and Dereli (2012) \cite{44} studied cosmological models considering the deceleration parameter to be linearly varying, where the Universe shows quintom like behaviour and ends with a big-rip. Lohakare et al. (2021) \cite{45} also investigated the cosmological model by assuming the deceleration parameter to be time varying in the context of $f(R,G)$ gravity, which shows quintessence behaviour at late times. Bakry and Shafeek (2019) \cite{46} also investigated the model of the Universe with time dependent deceleration parameter of the second degree and conclude that our Universe pass through a big-rip, then retreats back as it was initially in the instant of big bang. Tiwari et al. (2021) \cite{47} investigated time dependent deceleration parameter in the framework of $f(R,T)$ gravity and conclude that our Universe has a cyclic expansion history. That is the Universe starts with deceleration expansion and later shifts to accelerated expansion and further to super-exponential accelerating expansion period. Bishi et al. (2022) \cite{48} also investigated the behaviour of the FRLW cosmological model in the context of $f(R,T)$ gravity with quadratic deceleration parameter. Tiwari and Sofuoglu (2020) \cite{49} investigate quadratic varying deceleration parameter in the context of $f(R,T)$ gravity and establish an interesting result that Universe begins with a big bang and finally ends with a big rip. It also concludes that the Universe is filled with a quintessence like fluid in the early Universe and with a phantom like fluid at late time. Motivated by these studies and investigation, we present a new law with a varying deceleration parameter of third degree.\\

A brief introduction about FRW cosmological model with time dependent deceleration parameter is given in section \ref{sec1}. Modified field equation and solution of the model by introducing the deceleration parameter of third degree are discussed in section \ref{sec2} and \ref{sec3} respectively. Physical behaviour and the graphical representations of the cosmological parameters of the  proposed model are presented in section \ref{sec4}. We discussed the energy conditions and $Om(z)$ diagnostic parameter in section \ref{sec5} and \ref{sec6} respectively and the conclusion is given in section \ref{sec7}.\\

2. {\bf The field equations}:\label{sec2}
\setcounter {equation}{0}
\renewcommand {\theequation}{\arabic{equation}}\\
The Einstein field equations is given by

\begin{equation}\label{1}
G_{\mu \nu}\equiv R_{\mu \nu}-\frac{1}{2}Rg_{\mu \nu}=T_{\mu \nu}
\end{equation}
where $G_{\mu \nu}$ is the Einstein tensor, $R_{\mu \nu}$ is the Ricci tensor and $T_{\mu \nu}$ is the energy-momentum tensor.\\

Considering the homogeneous and isotropic space-time  Friedmann-Robertson-Walker (FRW) universe whose metric is given by
\begin{equation}\label{2}
ds^{2}=dt^{2}-a^{2}(t)\bigg[\frac{dr^{2}}{1-kr^{2}}+r^{2}(d\theta^{2}+sin^{2}\theta d\phi^{2})\bigg]
 \end{equation}
 where $a(t)$ is the scale factor and the spatial curvature index $k=-1$, $k=0$, and $k=1$ represents spatially open, flat and closed Universe respectively. We consider a co-moving fluid, where the FRW metric allows perfect fluid only for the energy-momentum tensor, which can be put in the form as follows
\begin{eqnarray}\label{3}
T_{\mu \nu}=(p+\rho)u_{\mu}u_{\nu}-pg_{\mu \nu}
\end{eqnarray}

where $\rho$ is the energy density and $p$ is the pressure, $g_{\mu \nu}$ is the fundamental Einstein tensor and $u_{\mu}$ is the four-velocity vector satisfying $u_{\mu}u^{\mu}=1$. For the metric \eqref{2}, the equation \eqref{1} takes the form

  \begin{equation}\label{4}
 \bigg(\frac{\dot{a}}{a}\bigg)^{2}=\frac{8\pi G}{3}\rho-\frac{k}{a^{2}}
\end{equation}

 \begin{equation}\label{5}
 \frac{\ddot{a}}{a}=\frac{-4\pi G}{3}(\rho+3p)
\end{equation}

Equation \eqref{4} gives the mathematical expression of energy density $(\rho)$ as
 \begin{equation}\label{6}
 \rho=\frac{3\dot{a}^{2}+3k}{8\pi Ga^{2}}
\end{equation}

 And equation \eqref{5} gives the mathematical expression of pressure $(p)$ as
 \begin{equation}\label{7}
 p=-\frac{1}{8\pi G}\bigg(\frac{2\ddot{a}}{a}+\frac{\dot{a}^{2}}{a^{2}}+\frac{k}{a^{2}}\bigg)
\end{equation}

  Equation of state of the model is given by
 \begin{equation}\label{8}
p=\omega\rho
\end{equation}
where $\omega$ is the equation of state parameter and $-1\leq \omega\leq 1$. \\

3. {\bf Time dependent deceleration parameter of the third degree}\label{sec3}:
\renewcommand {\theequation}{\arabic{equation}}\\

Berman (1983) \cite{39}, and Berman and Gomide (1988) \cite{40} put forward a new law of time varying Hubble parameter in the framework of Robertson-Walker space-time in general relativity that gives deceleration parameter, which is constant $q=m-1$, where $a$ is the scale factor and $m\geq0$ is a constant. Bakry and Shafeek (2019) \cite{46} also proposed a quadratic generalized varying deceleration parameter $q=(8n^{2}-1)-12nt+3t^{2}$, where $n$ is a constant greater than zero. In this work, we have introduced a time dependent deceleration parameter of third degree,

\begin{eqnarray}\label{9}
q=(6n^{3}-1)-22n^{2}t+18nt^{2}-4t^{3}
\end{eqnarray}
where $n$ is a positive constant. At $t=0$, $q=6n^{3}-1$ which is always positive for all $n\geq 1$, indicating that the expansion of the Universe starts with deceleration phase. For $0<n<1$, we have the value of the deceleration parameter to be negative which may be interpreted as inflation epoch just after Big Bang. The equation for the deceleration parameter is given by
\begin{eqnarray}\label{10}
q=\frac{-a\ddot{a}}{\dot{a}^{2}}=\frac{dH^{-1}}{dt}-1
\end{eqnarray}

From equation \eqref{9} and \eqref{10}, we get the Hubble parameter as

\begin{figure}\label{Fig.1}
\includegraphics[height=1.5in]{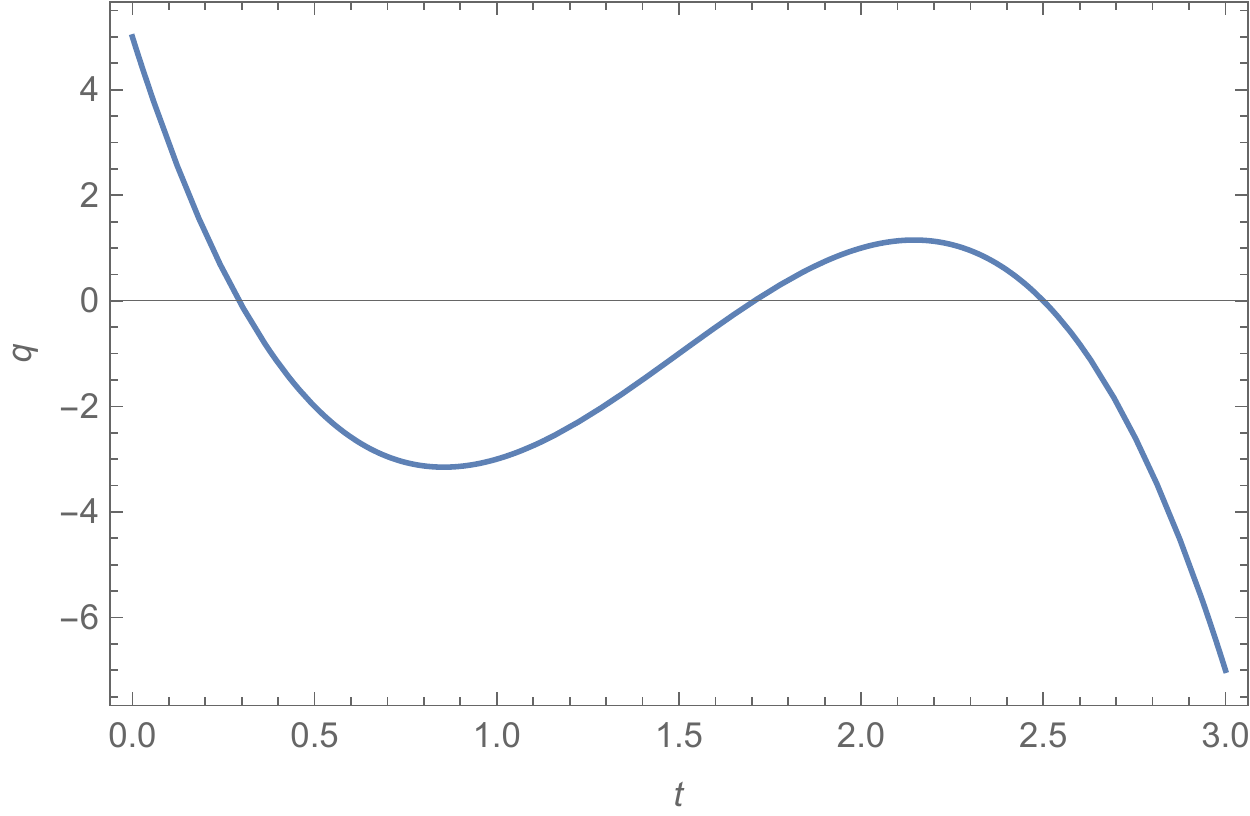}~~~~~~~~~~~
\includegraphics[height=1.5in]{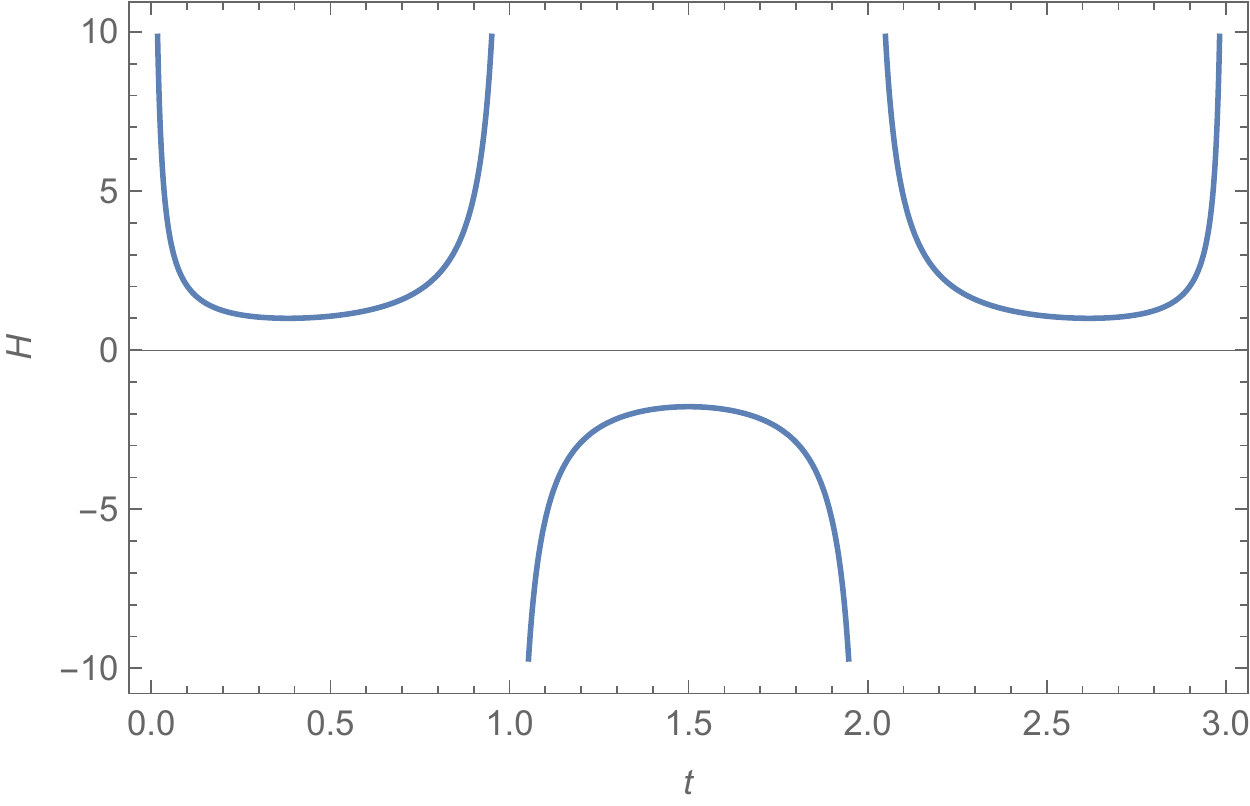}~~~~~\\

~~~~~~~Fig.1 Plot of $q$ vs. $t$ for $n=1$~~~~~~~~~~~~~~~~~~~~~Fig. 2 Plot of $H$ vs. $t$ for $n=1$

\end{figure}

\begin{eqnarray}\label{11}
H=\frac{\dot{a}}{a}=\frac{1}{t(n-t)(2n-t)(3n-t)}
\end{eqnarray}
From eq (11) we can see that Hubble parameter $H$ is always positive for $(n-t)<0$, $(2n-t)<0$ and $(3n-t)>0$ i.e Hubble parameter $H$ is positive in the range $0\leq t\leq3n$. Also we can observe that the Hubble parameter diverges at $t=0$, $t=n$, $t=2n$ and $t=3n$. Integrating equation \eqref{11}, we get
\begin{eqnarray}\label{12}
a=\bigg(\frac{t}{3n-t}\bigg)^{\frac{1}{6n^{3}}}\bigg(\frac{2n-t}{n-t}\bigg)^{\frac{1}{2n^{3}}}
\end{eqnarray}
which gives the expression of the scale factor $a$ as a function of cosmic time. It is observed that the scale factor $a=0$ for $t=0$ and $t=2n$, and it diverges for $t=n$ and $t=3n$, which shows that the Universe begin with big bang at $t=0$ and ends with a big crunch at $t=3n$.
 Putting the value of $a$, $\dot{a}$ and $\ddot{a}$ in the equation of $\rho$ and $p$ we get

\begin{figure}
\includegraphics[height=1.5in]{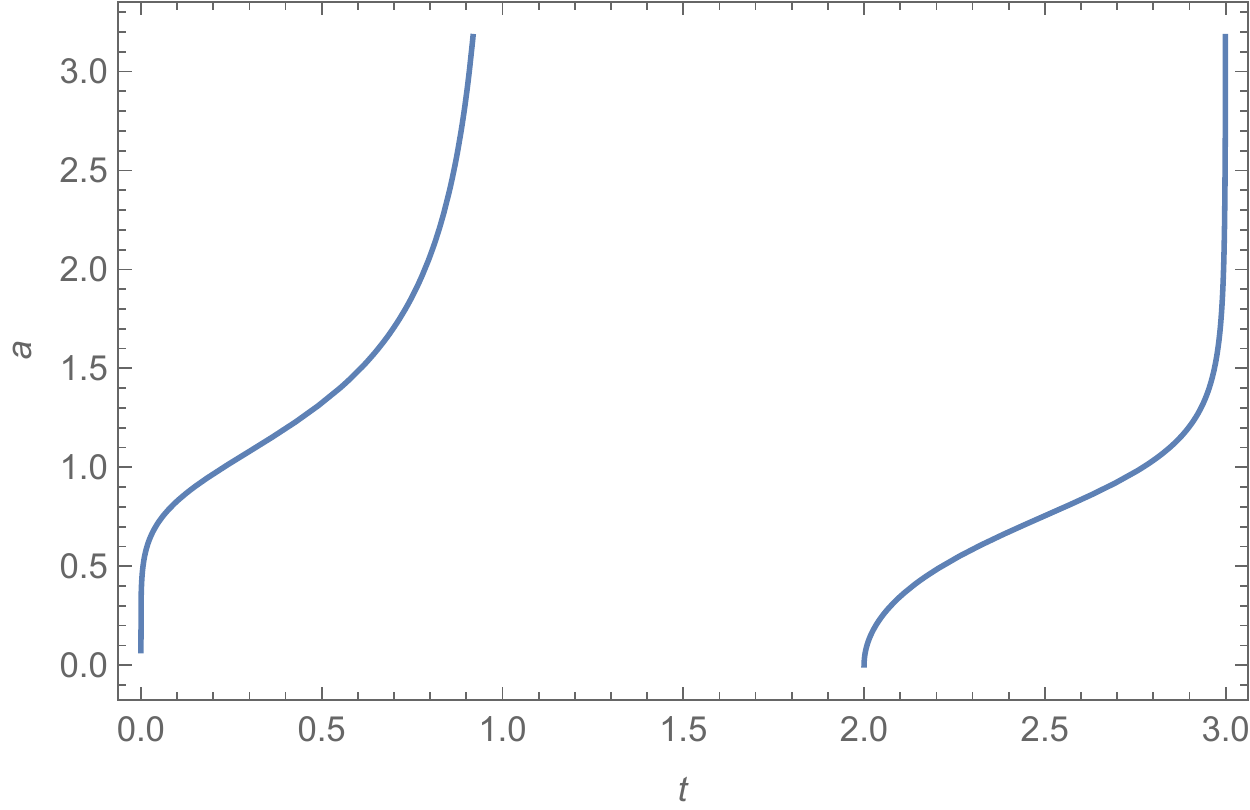}~~~~~~~~~~~
\includegraphics[height=1.5in]{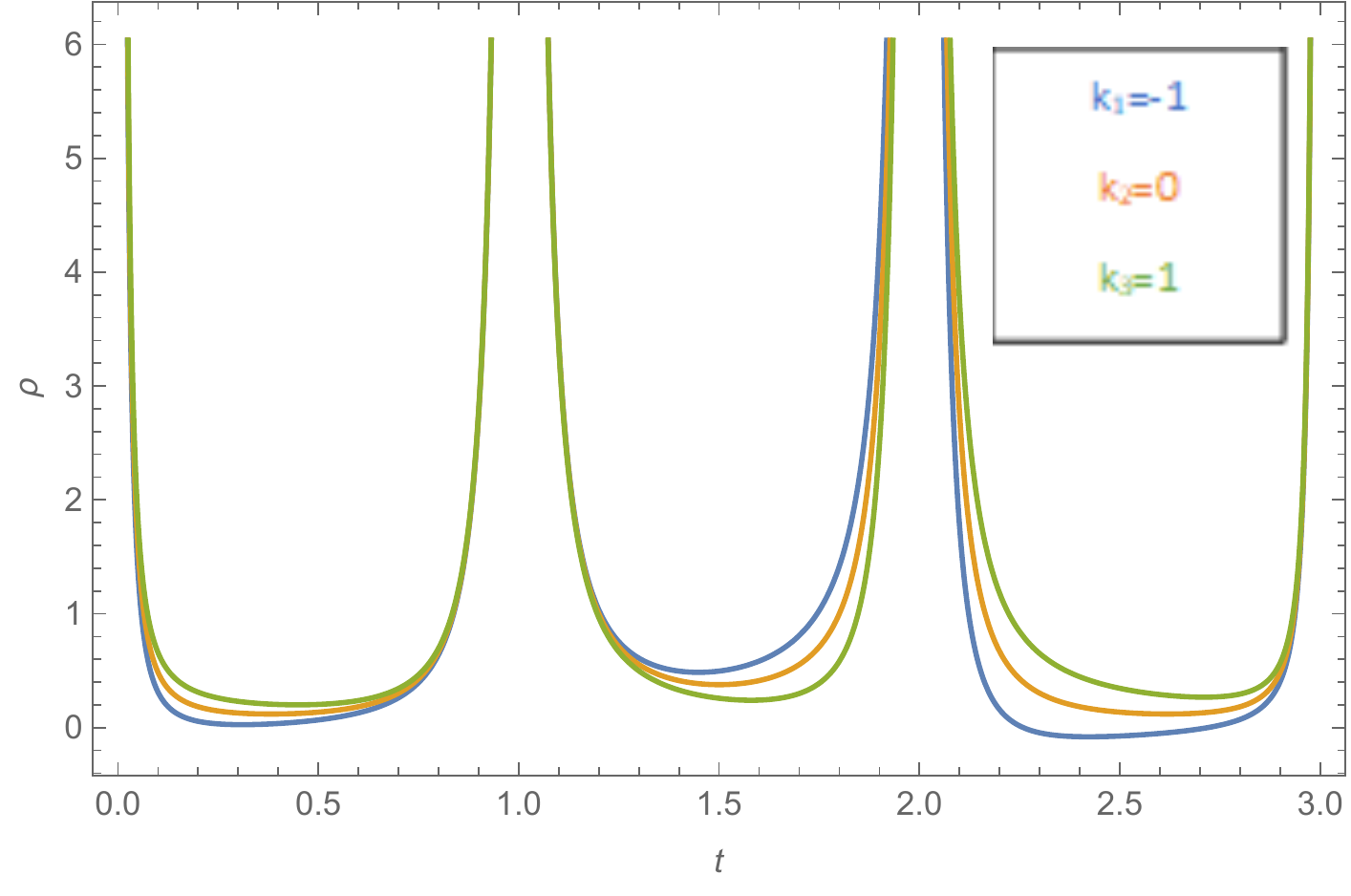}~~~~~\\

~~~~~~~Fig.3 Plot of $a$ vs. $t$ for $n=1$ ~~~~~~~~~~~~~~~Fig. 4 Plot of $\rho$ vs. $t$ for $n=1$ and $G=1$.

\end{figure}

\begin{eqnarray}\label{13}
 \rho=\frac{3}{8\pi G}\bigg[\frac{1}{t^{2}(n-t)^{2}(2n-t)^{2}(3n-t)^{2}}+k\bigg(\frac{3n-t}{t}\bigg)^{\frac{1}{3n^{3}}}\bigg(\frac{n-t}{2n-t}\bigg)^{\frac{1}{n^{3}}}\bigg]
\end{eqnarray}

\begin{eqnarray}\label{14}
 p&=&-\frac{1}{8\pi G}\bigg[\frac{3-12n^{3}+44n^{2}t+4t^{2}-36nt^{2}+4t^{3}}{t^{2}(n-t)^{2}(2n-t)^{2}(3n-t)^{2}}\cr
 &&+k\bigg(\frac{3n-t}{t}\bigg)^{\frac{1}{3n^{3}}}\bigg(\frac{n-t}{2n-t}\bigg)^{\frac{1}{n^{3}}}\bigg]
\end{eqnarray}

And equation of state parameter $\omega$ is given by

\begin{eqnarray}\label{15}
 \omega&=&-\frac{(3-12n^{3}+44n^{2}t+4t^{2}-36nt^{2}+4t^{3})t^{\frac{1}{3n^{3}}}(2n-t)^{\frac{1}{n^{3}}}}{3\bigg[t^{\frac{1}{3n^{3}}}(2n-t)^{\frac{1}{n^{3}}}+kt^{2}(n-t)^{\frac{2n^{3}+1}{n^{3}}}(2n-t)^{2}(3n-t)^{\frac{6n^{3}+1}{3n^{3}}}\bigg]}\cr
 &&-\frac{kt^{2}(n-t)^{\frac{2n^{3}+1}{n^{3}}}(2n-t)^{2}(3n-t)^{\frac{6n^{3}+1}{3n^{3}}}}{3\bigg[t^{\frac{1}{3n^{3}}}(2n-t)^{\frac{1}{n^{3}}}+kt^{2}(n-t)^{\frac{2n^{3}+1}{n^{3}}}(2n-t)^{2}(3n-t)^{\frac{6n^{3}+1}{3n^{3}}}\bigg]}
\end{eqnarray}
The commonly known examples of cosmological fluids with constant $\omega$ are dust ($\omega=0$), radiation ($\omega=\frac{1}{3}$) and vacuum energy ($\omega=-1$) which is also mathematically equivalent to cosmological constant $\Lambda$. When $\omega<-\frac{1}{3}$, it is considered in the context of dark energy as they gave rise to accelerated expansion of the Universe. Also $-1<\omega<-\frac{1}{3}$ indicates quintessence model and $\omega<-1$ represents phantom phase of the model. In our model, $\omega$ approaches to $-1$ as $t\rightarrow \infty$, this shows the accelerating expansion of the Universe at late epoch. The findings of Supernova Legacy Survey (Mark Sullivan, 2004) \cite{50} supports the dark energy model that can change into the epoch ($\omega<-1$)(Eisentein et al. 2005) \cite{51}, this observation is in agreement with our derived model. Since the  model pass in to dark energy epoch, then it comes to the phantom region $\omega<-1$, the existence of a big rip in future epoch is not ruled out by observational data and our model. \\

\begin{figure}
\includegraphics[height=1.5in]{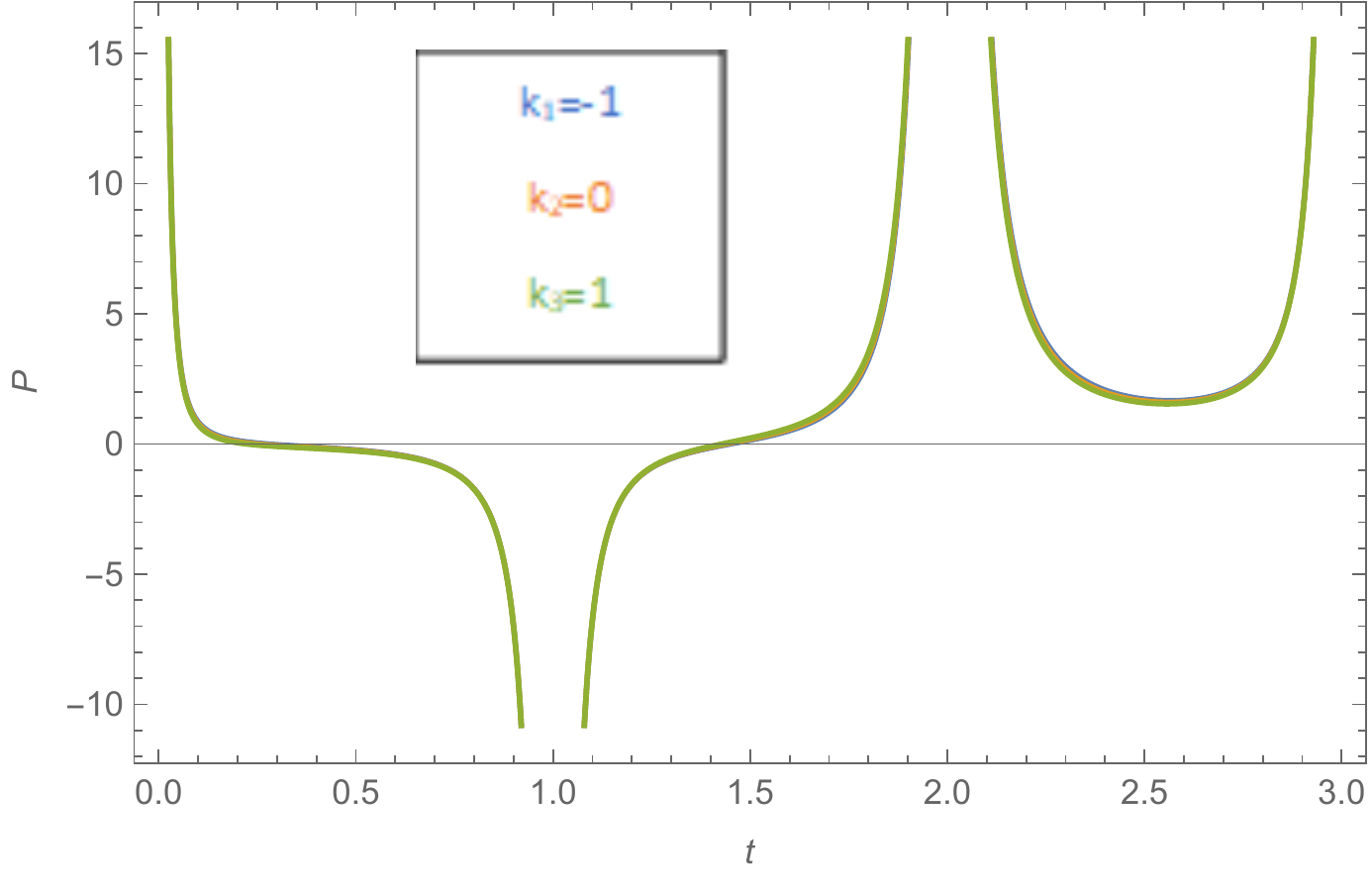}~~~~~~~~
\includegraphics[height=1.5in]{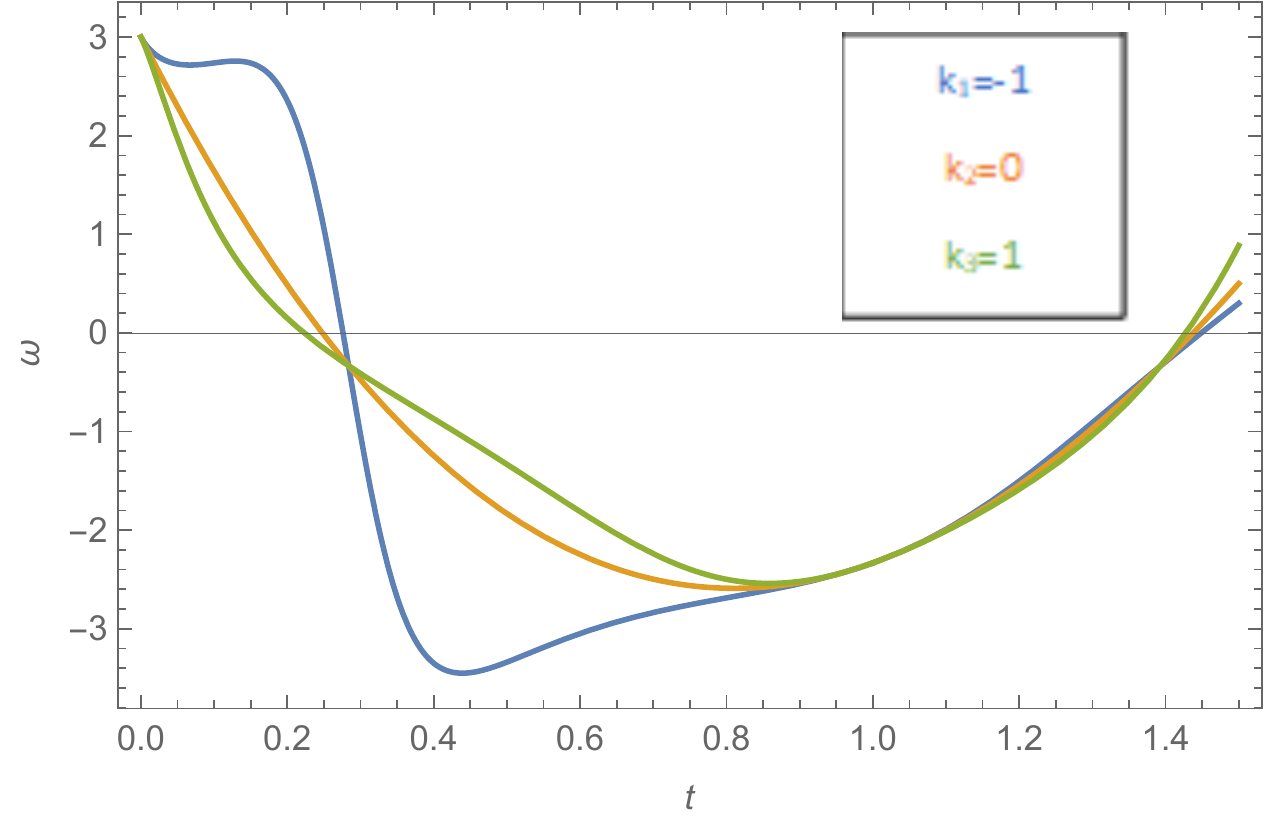}~~~~~~~~\\

~~Fig.5 Plot of $p$ vs. $t$ for $n=1$ and $G=1$. ~~~~~Fig.6 Plot of $\omega$ vs. $t$ for $n=1$ and $G=1$.
\end{figure}
\newpage
4. {\bf Physical behaviour of the cosmological parameters}\label{sec4}:
\renewcommand {\theequation}{\arabic{equation}}\\
For simplicity let us consider $n=1$ for our discussion. Fig. 1 shows the variation of the deceleration parameter $q$ w.r.t cosmic time $t$, which shows that the Universe starts with decelerating expansion $(q>0)$ and moves to an expansion phase with constant rate ($q=0$) and later transitioned to acceleration expansion phase $(-1\leq q<0)$ and to exponential acceleration phase (de Sitter expansion) $(q=-1)$ and further to super-exponential expansion phase ($q<-1$) and it retreats back again to decelerating expansion in periodic manner. Initially the deceleration parameter $q=5$, then enters into the acceleration phase $q<0$ at $t\in [0.29,1.7]$ and it enters into deceleration phase $q>0$ at $t\in[1.7,2.5]$ and ends at $t\in[2.5,3]$. This confirms the oscillating behaviour of our model (Broadhurst et al. 1990 \cite{52} and Morikawa 1990 \cite{53}). The model shows the accelerating expansion at two phases with the value of $q=-0.73$ at $t=0.35$ and $t=2.58$ which are in agreement with (Cunha and Lima 2008) \cite{54}. Comparing with the observational results of Cunha (2009) \cite{55}, the present value of deceleration parameter is $-0.73$ which gives the present age of Universe as $t=0.35$. Transitional phase of Universe from deceleration to acceleration is observed. From the plot of Hubble parameter $H$  in Fig. 2, we can say the Universe has singularity at $t=0$, big rip at $t= 1, 2$ and big-crunch at $t=3$. Fig. 3 shows the graphical representation of the scale factor $a$ w.r.t the cosmic time $t$. From the graph we can say that our model Universe has initial singularity at $t=0$ as $a(t)=0$. Then $a(t)$ diverges at $t=1$, which gives the big rip behaviour of the universe (Caldwell et al. 2003) \cite{56} and later retreats into moment of singularity at $t=2$ and again diverges at $t=3$ again observing big rip behaviour (Caldwell et al. 2003), which shows a periodic nature of the Universe predicting a periodic Universe. From Fig. 4 we can say that the positive energy density condition for spatially open, flat, and closed Universe holds true. Further we can also say that the Universe has initial singularity at $t=0$, big rip at $t= 1, 2$ and big-crunch at $t=3$ (Caldwell et al. 2003) \cite{56}. Fig. 5 highlights the graph of pressure w.r.t cosmic time $t$, which also shows that the Universe has initial singularity at $t=0$, big rip at $t= 1, 2$ and big-crunch at $t=3$ for spatially open, flat, and closed Universe. From the plot we know that the positivity condition of pressure is violated for spatially open, flat, and closed Universe, which reckons for the accelerating expansion of the Universe. Fig. 6 shows the evolution of the equation of the state parameter (EOS) $\omega$ with respect to cosmic time $t$ for spatially open, flat, and closed Universe. It manifest simple different behaviour for spatially open, flat, and closed Universe but behaves in a similar manner, which starts from positive value (stiff matter dominated $\omega=1$ and $\omega=1/3$ and tends zero (dust filled model, $\omega=0$) and further move towards negative value (vacuum energy model $\omega=-1$ and phantom energy $\omega<-1$). The solutions of this model with varying deceleration parameter of cubic degree indicates that Universe begins with inflation for few moments after the big bang then it decelerates and accelerated until the moment of big rip, then further accelerates to the moment of future crunch.\\

5. {\bf Energy conditions}\label{sec5}:
\renewcommand {\theequation}{\arabic{equation}}\\

Energy conditions are a vital and important tools to know and investigate the geodesics of the Universe. It is primarily the boundary conditions to maintain the positivity condition of the energy density (Hawking and Ellis 1973 \cite{57}, Poisson 2004 \cite{58}). These energy conditions do not corresponds to the physical reality. The observable effects of dark energy which is manifested in the violation of strong energy condition is the most recent examples of this reality. The four basic/fundamental energy conditions are. \\
(i) Strong energy condition : $\Rightarrow$ $\rho+3p\geq0$.\\
(ii) Null energy condition : $\Rightarrow$ $\rho+p\geq0$.\\
(iii) Weak energy condition : $\Rightarrow$ $\rho\geq0$, $\rho+p\geq0$.\\
(iv) Dominant energy condition : $\Rightarrow$ $\rho\geq0$, $|p|\leq\rho$.\\

\begin{figure}
\includegraphics[height=1.5in]{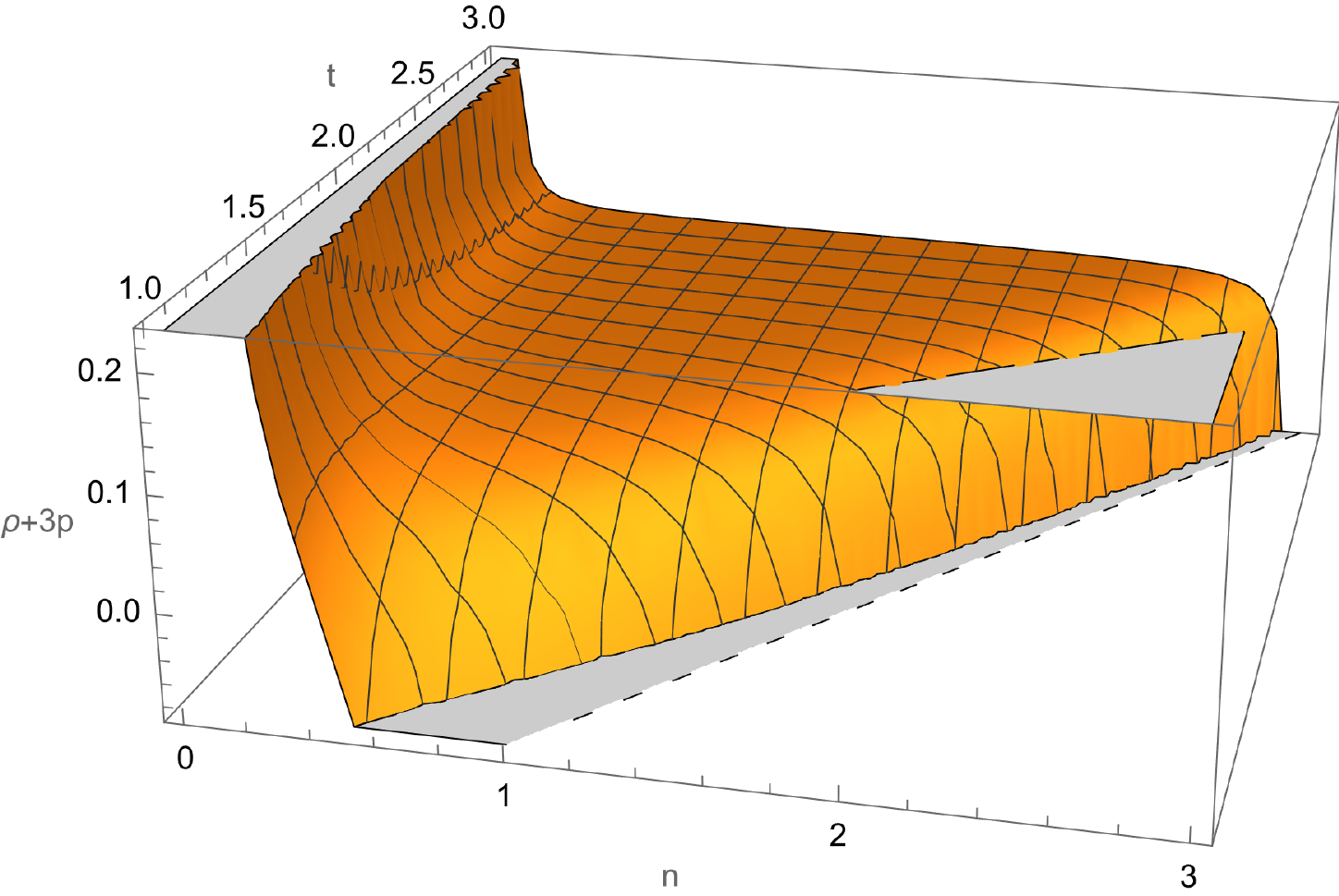}~~~~~
\includegraphics[height=1.5in]{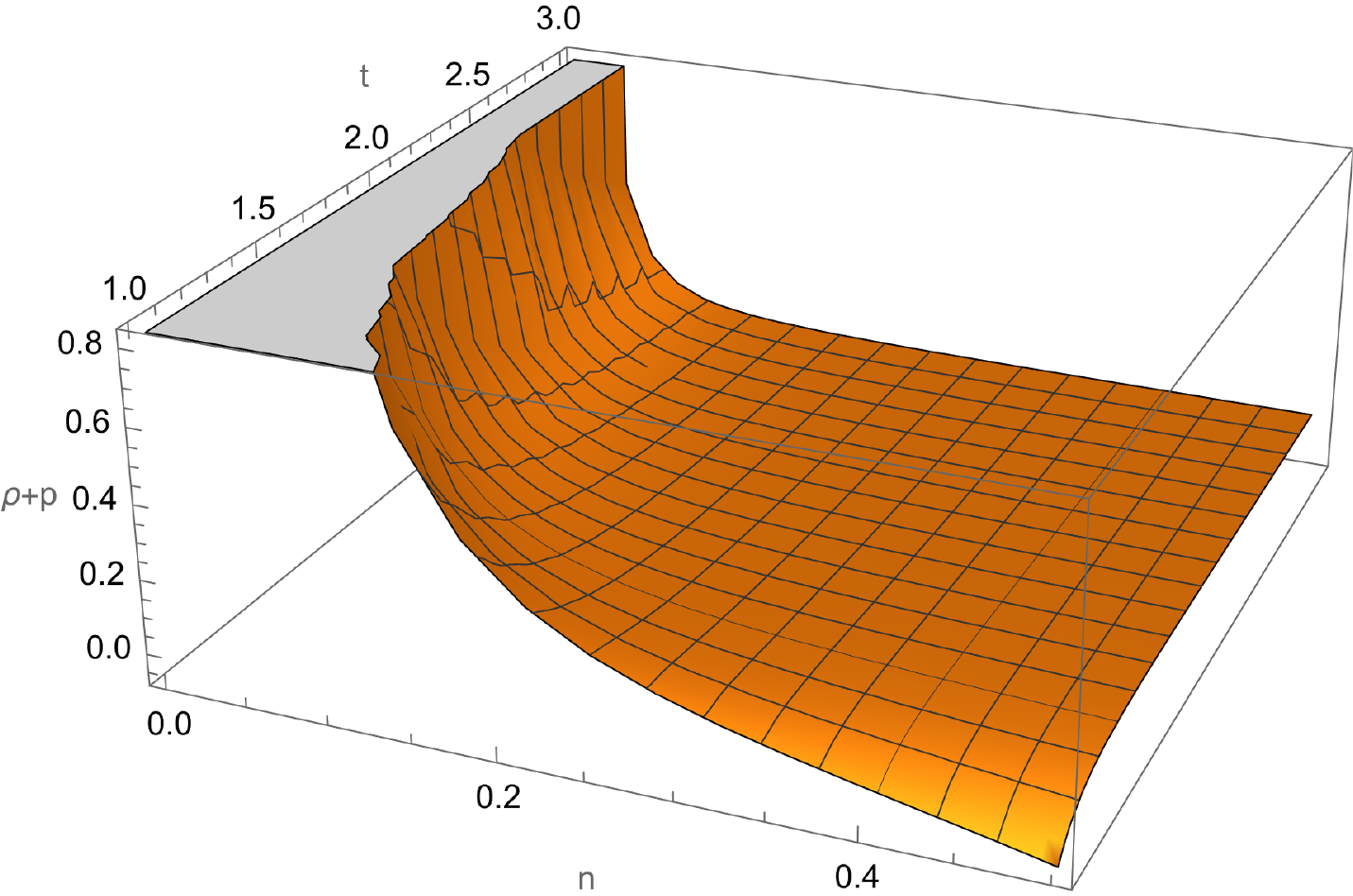}~~~~~\\

Fig.7 3D Plot of SEC for $k=1$, $n=1$ and $G=1$.~~Fig.8 3D Plot of NEC for $k=1$, $n=1$ and $G=1$.

\end{figure}
\begin{figure}
\includegraphics[height=1.5in]{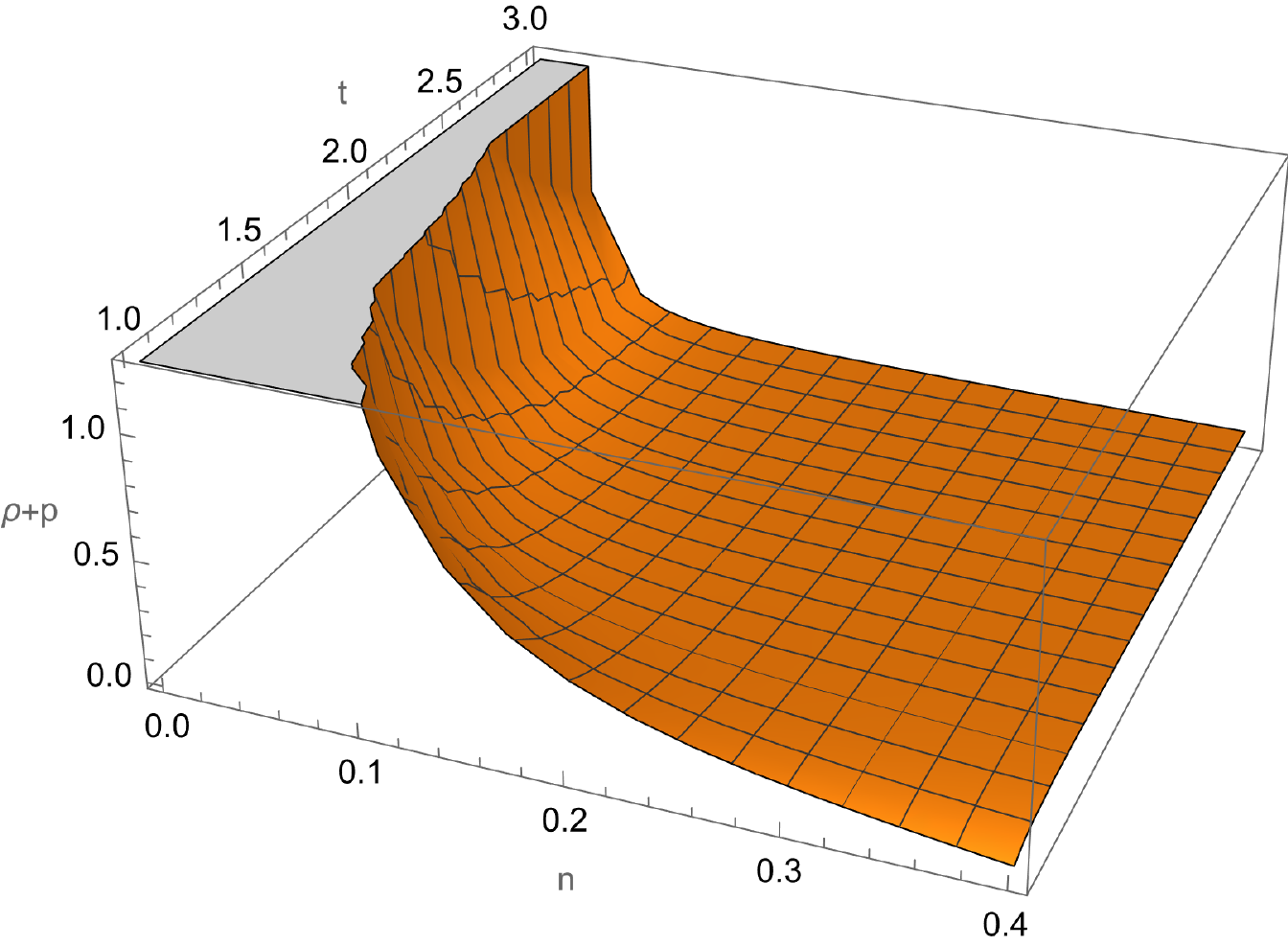}~~~~~
\includegraphics[height=1.5in]{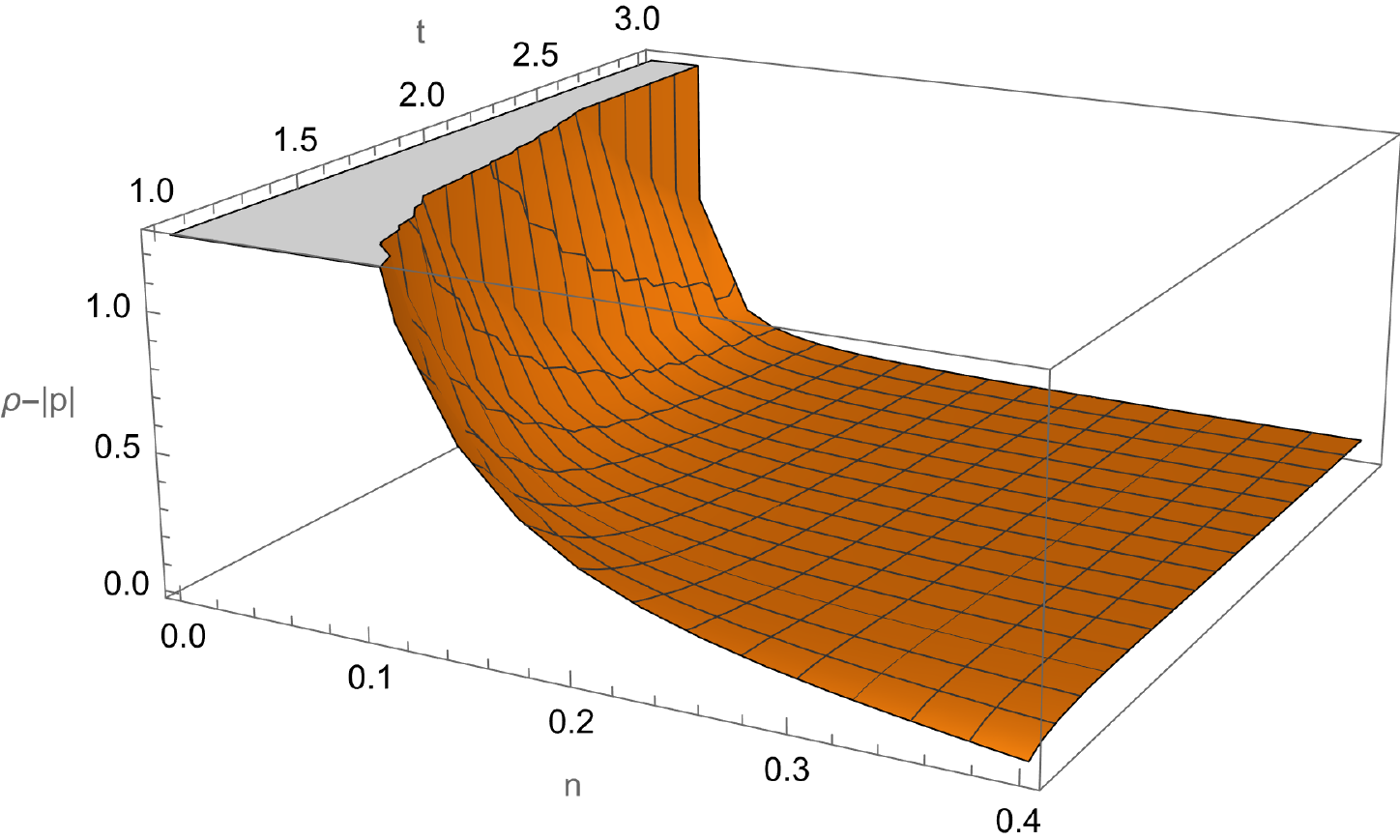}~~~~~\\

Fig.9 3D Plot of WEC for $k=1$, $n=1$ and $G=1$.~~Fig.10 3D Plot of DEC for $k=1$, $n=1$ and $G=1$.
\end{figure}

Substituting the values of equation \eqref{13} and \eqref{14} in the energy conditions we get\\
SEC:
\begin{eqnarray}\label{16}
 &&\frac{3}{8\pi G}\bigg[-\frac{3-12n^{3}+44n^{2}t+4t^{2}-36nt^{2}+4t^{3}}{t^{2}(n-t)^{2}(2n-t)^{2}(3n-t)^{2}}+k\bigg(\frac{3n-t}{t}\bigg)^{\frac{1}{3n^{3}}}\bigg(\frac{n-t}{2n-t}\bigg)^{\frac{1}{n^{3}}}\cr
&& +\frac{1}{t^{2}(n-t)^{2}(2n-t)^{2}(3n-t)^{2}}+k\bigg(\frac{3n-t}{t}\bigg)^{\frac{1}{3n^{3}}}\bigg(\frac{n-t}{2n-t}\bigg)^{\frac{1}{n^{3}}}\bigg]
   \geq0
  \end{eqnarray}
WEC:
\begin{eqnarray}\label{17}
\frac{3}{8\pi G}\bigg[\frac{1}{t^{2}(n-t)^{2}(2n-t)^{2}(3n-t)^{2}}+k\bigg(\frac{3n-t}{t}\bigg)^{\frac{1}{3n^{3}}}\bigg(\frac{n-t}{2n-t}\bigg)^{\frac{1}{n^{3}}}\bigg]\geq0
  \end{eqnarray}

\begin{eqnarray}\label{18}
    &&-\frac{1}{8\pi G}\bigg[\frac{3-12n^{3}+44n^{2}t+4t^{2}-36nt^{2}+4t^{3}}{t^{2}(n-t)^{2}(2n-t)^{2}(3n-t)^{2}}+k\bigg(\frac{3n-t}{t}\bigg)^{\frac{1}{3n^{3}}}\bigg(\frac{n-t}{2n-t}\bigg)^{\frac{1}{n^{3}}}\bigg]\cr
    &&+\frac{3}{8\pi G}\bigg[\frac{1}{t^{2}(n-t)^{2}(2n-t)^{2}(3n-t)^{2}}+k\bigg(\frac{3n-t}{t}\bigg)^{\frac{1}{3n^{3}}}\bigg(\frac{n-t}{2n-t}\bigg)^{\frac{1}{n^{3}}}\bigg]\geq0
  \end{eqnarray}

NEC:
\begin{eqnarray}\label{19}
    &&-\frac{1}{8\pi G}\bigg[\frac{3-12n^{3}+44n^{2}t+4t^{2}-36nt^{2}+4t^{3}}{t^{2}(n-t)^{2}(2n-t)^{2}(3n-t)^{2}}+k\bigg(\frac{3n-t}{t}\bigg)^{\frac{1}{3n^{3}}}\bigg(\frac{n-t}{2n-t}\bigg)^{\frac{1}{n^{3}}}\bigg]\cr
    &&+\frac{3}{8\pi G}\bigg[\frac{1}{t^{2}(n-t)^{2}(2n-t)^{2}(3n-t)^{2}}+k\bigg(\frac{3n-t}{t}\bigg)^{\frac{1}{3n^{3}}}\bigg(\frac{n-t}{2n-t}\bigg)^{\frac{1}{n^{3}}}\bigg]\geq0
  \end{eqnarray}

 DEC:
\begin{eqnarray}\label{20}
\frac{3}{8\pi G}\bigg[\frac{1}{t^{2}(n-t)^{2}(2n-t)^{2}(3n-t)^{2}}+k\bigg(\frac{3n-t}{t}\bigg)^{\frac{1}{3n^{3}}}\bigg(\frac{n-t}{2n-t}\bigg)^{\frac{1}{n^{3}}}\bigg]\geq0
  \end{eqnarray}

  \begin{eqnarray}\label{21}
   &&\bigg|-\frac{1}{8\pi G}\bigg[\frac{3-12n^{3}+44n^{2}t+4t^{2}-36nt^{2}+4t^{3}}{t^{2}(n-t)^{2}(2n-t)^{2}(3n-t)^{2}}+k\bigg(\frac{3n-t}{t}\bigg)^{\frac{1}{3n^{3}}}\bigg(\frac{n-t}{2n-t}\bigg)^{\frac{1}{n^{3}}}\bigg]\bigg|\cr
   &&\leq\frac{3}{8\pi G}\bigg[\frac{1}{t^{2}(n-t)^{2}(2n-t)^{2}(3n-t)^{2}}+k\bigg(\frac{3n-t}{t}\bigg)^{\frac{1}{3n^{3}}}\bigg(\frac{n-t}{2n-t}\bigg)^{\frac{1}{n^{3}}}\bigg]
  \end{eqnarray}\\

The 3D graph of the strong energy condition is given in Fig. 7. From the graph we see that the strong energy condition does not hold true for our model Universe, which agrees with the present observation. Fig. 8, 9 and 10 shows the 3D plot of the weak energy condition, null energy condition, and dominant energy condition respectively. We see from the graph that the null energy condition, weak energy condition, and dominant energy conditions holds true in our model Universe.\\

6. {\bf $Om(z)$ diagnostic parameter}\label{sec6}:
\renewcommand {\theequation}{\arabic{equation}}\\
The relation between the scale factor and redshift is given by
\begin{eqnarray}\label{22}
  a(t)=\frac{1}{1+z}
\end{eqnarray}
This relation further relates cosmic time and redshift as
\begin{eqnarray}\label{23}
  \frac{1}{(1+z)^{6}+1} &=& \frac{t^{6}}{2}-\frac{9}{4}t^{2}+\frac{21}{8}t-\frac{t}{8(3-2t)}\cr
  \Rightarrow\frac{t^{3}}{2} &\simeq& \frac{1}{(1+z)^{6}+1}\cr
  \Rightarrow t &\simeq& \big[\frac{2}{(1+z)^{6}+1}\big]^{\frac{1}{3}}
\end{eqnarray}

$Om(z)$ diagnostic parameter is an important geometrical diagnostic approach used in literature. Literature shows that the dark energy models are associated with positive Hubble parameter and negative deceleration parameter. $H$ and $q$ alone are not enough to differentiate different dark energy models. To investigate such analysis $Om(z)$ diagnostic parameter is presented. When $Om(z)$ is constant w.r.t red-shift, the dark energy model is in the form of $\Lambda$. The nature of the gradient of $Om(z)$ characterize the dark energy models in this manner: when the gradient of the $Om(z)$ is positive it indicates phantom phase $\omega<-1$, and when the gradient of the $Om(z)$ is negative it indicates quintessence phase $\omega>-1$ (Sahni et al. 2008) \cite{59}.

The $Om(z)$ diagnostic parameter can be defined as

\begin{eqnarray}\label{24}
Om(z)=\frac{\bigg[\frac{H(z)}{H_{0}}\bigg]^{2}-1}{(1+z)^{3}-1}
\end{eqnarray}
Here $H_{0}$ denotes the present value of the Hubble parameter.\\

\begin{figure}
~~~~~~~~~~~~~~~~~~~~~~~~~~~~~~~~~~~\includegraphics[height=1.5in]{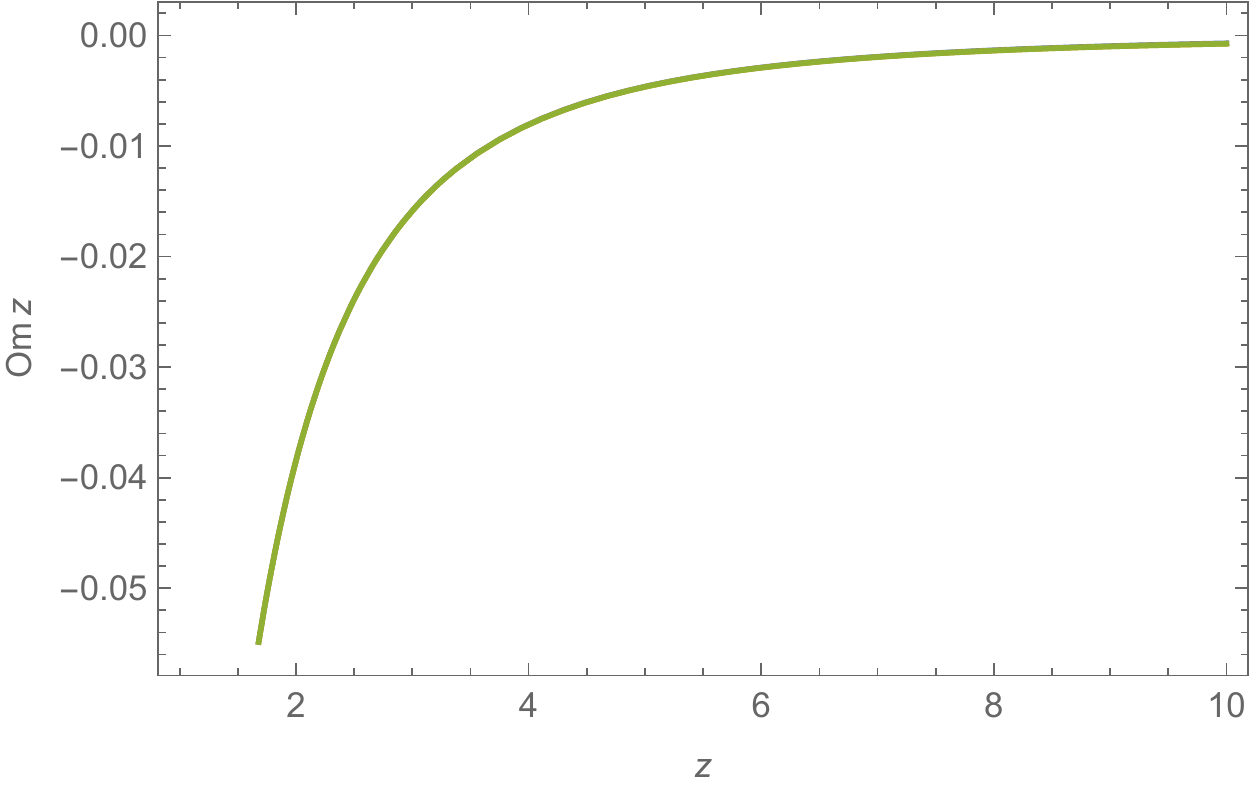}~~~~~

~~~~~~~~~~~~~~~~~~~~~~~~~~~~~Fig.11 Plot of $Om(z)$ vs. $z$ for $n=1$ and $H_{0}=67.77$.
\end{figure}

Fig. 11 represents the plot of $Om(z)$ diagnostic parameter versus red-shift $z$ for $n=1$ and $H_{0}=67.77$. From Fig. 11 we observe that slope of $Om(z)$ is negative which indicates phantom nature ($\omega<-1$) of our Universe.

7. {\bf Conclusion}\label{sec7}:
\renewcommand {\theequation}{\arabic{equation}}\\
In this work we have investigated FRW cosmology with variable deceleration parameter of third degree of the form $q=(6n^{3}-1)-22n^{2}t+18nt^{2}-4t^{3}$. At initial moment, Universe shows the inflation for $n<1$ then it decelerates for $n\geq1$. In this model, we also observe the transition phase from deceleration to acceleration for the valid range $n\geq 1$ which is in agreement with observations. From the behavior of the cosmological parameters discussed above, we see that the parameters $a$, $H$, $q$, $\rho$ and $p$ indicates that our model has a point-type singularity initially at $t=0$ and similar singularities occur periodically at $t=n, 2n$ and observes future crunch at late time $t=3n$. We also see that the parameters $a$, $H$, $q$, $\rho$, and $p$ are infinite initially and preserve the periodic behaviour against cosmic time. Thus our model exhibits a periodic Universe. Also the positivity condition of energy density $\rho$ is maintained and satisfied in our model. Strong energy condition is violated due to the negative nature of the pressure, accounting for the accelerated expansion of the Universe. But weak energy condition, null energy condition and dominant energy condition are satisfied. The slope of $Om(z)$ diagnostic parameter with respect to red-shift $z$ is negative indicating the phantom phase of our model Universe. Thus the model presented in this paper can be helpful in the understanding of the evolution of the Universe.\\

\section*{Declaration of conflict of interest} The author declares no potential conflict of interest or competing interest.

\end{document}